%% file: paper.tex
\documentclass[a4paper,10pt]{article}
\usepackage[utf8]{inputenc}

\pdfoutput=1 

\usepackage{graphicx}
\usepackage{subfigure}
\usepackage{hyperref}

\title{Open Tiled Manycore System-on-Chip\thanks{Dagstuhl Seminar
    13052, ``Multicore Enablement for Embedded and Cyber Physical
    Systems'', organizers: Andreas Herkersdorf, Michael G. Hinchey,
    Michael Paulitsch, 27.01.13--01.02.13}}

\author{Stefan Wallentowitz \and Philipp Wagner \and Michael
  Tempelmeier \and Thomas Wild \and Andreas Herkersdorf \medskip \\
  Institute for Integrated Systems, TU München \\ Arcisstr. 21,
  D--80290 München \\ \small{\texttt{stefan.wallentowitz@tum.de}}}

\begin{document}

\maketitle

\begin{abstract}
  Manycore System-on-Chip include an increasing amount of processing
  elements and have become an important research topic for
  improvements of both hardware and software. While research can be
  conducted using system simulators, prototyping requires a variety of
  components and is very time consuming. With the \emph{Open Tiled
    Manycore System-on-Chip~(OpTiMSoC)} we aim at building such an
  environment for use in our and other research projects as
  prototyping platform.

  This paper describes the project goals and aspects of OpTiMSoC and
  summarizes the current status and ideas.
\end{abstract}

\input{motivation}
\input{basicelements}

\input{targets}

\input{pgen}
\input{programming}
\input{debugging}
\input{roadmap}

\bibliographystyle{plain}
\bibliography{paper}

\end{document}

%% file: motivation.tex
\section{Motivation \& Project Goals}

Multicore System-on-Chip have become dominant in research and industry
as high-performance and power-efficient processing platforms. With
increasing amounts of processing elements the tile organization is a
popular way to organize the elements of such a platform. Those
platforms base on Network-on-Chip with a regular organization, often a
mesh. Processing elements, memories and I/O devices are connected to
this interconnect as \emph{tiles}. This allows for platform creation
by replicating these basic tiles to larger platforms. Some examples
for tiled platforms are Tilera's processors or Intel's ``Single Chip
Cloud Computer''.

Many aspects of such platforms are currently in the focus of research,
for example system design aspects, communication-to-computation
coupling, coherency and consistency issues, programming of future
massively parallel platforms and many more.

\medskip

Research of improvements of future manycore system-on-chip is
mainly performed using system simulation approaches, where a variety
of tools exist. Nevertheless when it comes to prototyping, hardware
architects rely on a platform they can prototype their ideas with.
Building such a platform from scratch is very time consuming and a
complete -- not even speaking of common -- platform has not been
established. OpTiMSoC tries to fill this gap by providing the
necessary environment to augment research prototyping in the
respective field.

\medskip

The goals of OpTiMSoC can therefore be summarized as follows:

\begin{itemize}
\item Build a foundation library of hardware elements that can shape a
  variety of tiled manycore platforms
\item Support several different target platforms, ranging from
  simulation to FPGA-based emulation
\item Provide the required infrastructure to compose and build those
  platforms
\item Include a programming environment and runtime system
\item Enable debugging with trace-based debugging techniques with a
  common middleware layer
\end{itemize}

Along those goals this paper will present the current status of the
OpTiMSoC project. The current development of the project can be found
on the project website\footnote{\url{http://www.optimsoc.org}}. 
The project is open to contributions and aims to always provide paths that do
not require licenses for necessary tools.

%% file: basicelements.tex
\section{Basic Hardware Elements}

The central element of OpTiMSoC is
LISNoC\footnote{\url{http://www.lisnoc.org}}, a basic
\textbf{Network-on-Chip} implementation we developed. The basic NoC is
a packet-switched, wormhole-forwarding, buffered implementation
supporting virtual channels to avoid message dependent deadlocks.
Variations supporting priorities, multicasts or bufferless forwarding
are also part of LISNoC.

The Network-on-Chip is the underlying structure for the whole
platform. Mesh topologies are commonly deployed in tiled manycore
System-on-Chip, but also other topologies such as rings or
hierarchical structures can be easily implemented using the LISNoC
hardware elements library.

\medskip

Beside the communication foundation, the \textbf{processing elements}
are the important elements in OpTiMSoC. The basic processing element
used in OpTiMSoC is the OpenRISC
processor\footnote{\url{http://www.opencores.org/openrisc}}. Despite
the current infrastructure is built around this processor core, other
alternatives will be added in the future, for example the LEON3 Sparc
implementation\footnote{\url{http://www.gaisler.com/}}.

Some peripheral elements for the tiles, such as memories, interconnect
etc. are also part of the library, together with additional I/O
elements. Those strongly depend on the target platform as described in
the following.

\medskip

Crucial elements, which are under research currently, are those
bridging communication and computation, namely the \textbf{network
  adapter} or also network interface (NA/NI). The central elements of
such network adapters are:

\begin{itemize}
\item Handle \emph{memory transfers} between tiles and the memories.
\item Provide hardware means to send \emph{messages} between the tiles
  such as for processor communication or external data streams.
\end{itemize}

\begin{figure}
  \centering
  \subfigure[Distributed memory]{
    \includegraphics[scale=.9]{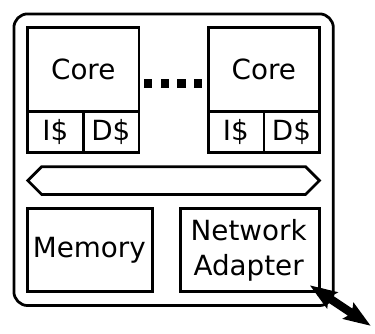}
    \label{fig:tile_dm}
  }
  \subfigure[Partitioned global address space]{
    \includegraphics[scale=.9]{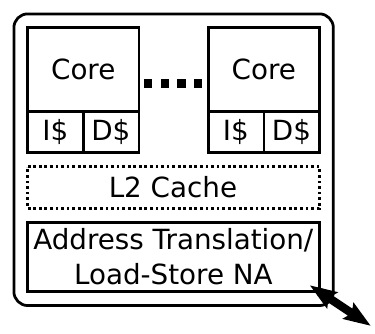}
    \label{fig:tile_pgas}
  }
  \subfigure[Shared memory]{
    \includegraphics[scale=.9]{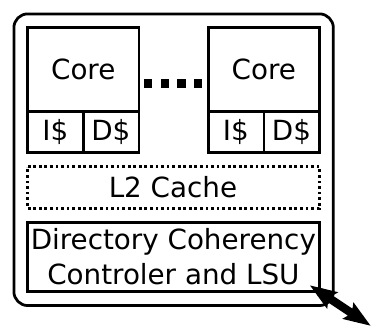}
    \label{fig:tile_sm}
  }
  \caption{Different tile implementation alternatives}
  \label{fig:tiles}
\end{figure}

\medskip

The services provided by a network adapter implementation depend on
the system and \textbf{tile organization}. In OpTiMSoC we aim to cover
three central organization styles as sketched in Figure~\ref{fig:tiles}.
The choice of the organization strongly correlates with the choice of
the programming model. At the moment we focus on distributed memory
organization and message passing programming.

In distributed memory (see Figure~\ref{fig:tile_dm}) a varying amount
of processor cores is connected to a locally shared memory. Due to
restricted memory sizes, a partitioned global address space variant
(see Figure~\ref{fig:tile_pgas}) is used in prototyping platforms,
where the network adapter employs an MPU-like load-store unit (LSU)
that translates memory access addresses in a way that the global
memory is partitioned in separate chunks, each assigned to one tile.

Finally, we currently work towards a platform variant which provides
globally shared memory among all tiles (see Figure~\ref{fig:tile_sm}).
In the other platforms the locally shared memory or address space
needs to be coherent, which is done with a write-through snooping
policy. In future systems it is planned to augment this with a level 2
write-back directory-based cache coherency.

%% file: targets.tex
\section{Target Platforms}

With the current releases we try to cover a basic number of target
platforms for development and prototyping of manycore System-on-Chip.
In the following we describe the basic ways to implement an OpTiMSoC
instance currently as they cover the main use cases.

\begin{description}
\item[RTL simulation using EDA tools] is the basic entry point from a
  hardware designers point of view. For development of hardware
  elements in OpTiMSoC this step is essential. Nevertheless, most
  tools are commercial, so that a second variant (Verilator) is
  preferred in other cases.
\item[Verilated simulation] is a variant of RTL simulation.
  Verilator\footnote{\url{http://www.veripool.org/wiki/verilator}} is
  a toolchain that compiles RTL code to C++ code for simulation. A
  SystemC\footnote{\url{http://www.accelera.org}} module can easily be
  generated by Verilator. The main advantage of the use of Verilator
  over RTL simulation with (potentially costly) EDA tools is that it
  is completely open source. That allows software developers to
  develop code without prototyping hardware or the necessity for a
  commercial simulation tool.
\item[Xilinx University Program Board 5
  (XUPV5)]\footnote{\url{http://www.xilinx.com/univ/xupv5-lx110t.htm}}
  is a widely used FPGA board often found in academic institutions.
  It employs a mid-range Virtex 5 FPGA and multiple I/O. Currently the
  DDR memory, UART and LCD display are used in OpTiMSoC.
\item[ZTEX 1.15 boards]\footnote{\url{http://www.ztex.de}} are small
  outline boards including a Spartan 6 FPGA, DDR3 memory and a Cypress
  EZ-USB interface chip. The boards are easily used in standalone mode
  only, using a USB cable. Two small variants do not require a costly
  synthesis license but can be used with the XILINX WebPack. The
  boards are especially powerful using our debug infrastructure as
  described below and can be very useful to software developers.
\item[CHIPit emulation
  platform]\footnote{\url{http://www.synopsys.com}} is an emulation
  platform based on multiple FPGAs and allows for the prototyping of
  much larger systems. In the future OpTiMSoC will include a setup for
  such large systems on the emulator platform and the releated
  infrastructure.
\end{description}

The target platforms described above are of course only a starting
point. We aim to support many more targets in the future. Once the
basic operation and I/O of other targets is supported (or is already
worked out from other uses) it is relatively simple to include them in
OpTiMSoC.

%% file: pgen.tex
\section{Platform Generator Tool}

As introduced before, OpTiMSoC is not a platform proposal for one
specific setup regarding dimensions or organization of the system.
Instead it is intended as a basic set of elements and the necessary
infrastructure. To allow the user to generate many different variants
and configuration for a variety of targets easily, we previously
presented our envisioned tool flow in~\cite{FPL}.

The elements, tile organization and target platforms described above
can be seen as the library elements to a platform generator tool.
Based on these elements two separate platform generation steps are
envisioned:

\begin{description}
\item[Platform description mapping:] The platform is described based
  on generic layout patterns. It defines the interconnect structure,
  the organization of tiles in the system and the tile-internal
  organization. This description is augmented with some basic
  parameters, such as the processor implementation or similar. The
  output is an OpTiMSoC configuration.
\item[OpTiMSoC configuration mapping:] An OpTiMSoC configuration can
  be either described manually by the designer or can be the output of
  the platform description mapping. This mapping generates the actual
  files for the used elements for a given target and the respective
  build files, such as scripts or makefiles.
\end{description}

For a more detailed discussion we refer the reader to~\cite{FPL}.

%% file: programming.tex
\section{Programming and Runtime System}

Previously we presented the hardware elements and targets of OpTiMSoC.
Of same importance of course is the software part of OpTiMSoC, what
involves the question how to program it and what runtime system it
employs. Basically, two \textbf{underlying systems} are the
fundamental base of OpTiMSoC software:

\begin{description}
\item[Baremetal system] The baremetal system involves all drivers
  necessary to execute on the processing hardware, which involves the
  stack and heap management, exception handling etc. Therefore we
  essentially provide a port of the newlib libc implementation for
  OpTiMSoC based on the OpenRISC development (that also includes the
  gcc compiler).
\item[Lean runtime system] Based on the runtime system a simple
  runtime system is implemented that provides the central functions
  needed for more sophisticated systems: a thread scheduler and
  virtual memory management. Based on these, a simple microkernel or
  even other operating systems can be ported. It has to be noted that
  the runtime system, as most other elements of OpTiMSoC, is not
  optimized in a sense that we tried to get every cycle out of there,
  but was instead designed with the goals of readability and
  simpleness.
\end{description}

Both systems are augmented with the basic drivers for the hardware
elements. Furthermore high-level \textbf{programming APIs} are
required for parallel programming. We decided to provide baseline
implementations of the Multicore Association
APIs\footnote{\url{http://www.mutlicore-association.org}} as they
perfectly cover the problems of embedded systems. Therefore we
currently work on the implementation of:

\begin{description}
\item[MCAPI] The communication API is a basic message passing library
  used for inter-process communication. The API handles the
  communication over the Network-on-Chip abstracting from the employed
  network adapter capabilities. Different transport layer
  implementations instead handle the different hardware implementations.
\item[MTAPI] Recently the task management API was released. We work
  towards a distributed implementation of the MTAPI in OpTiMSoC. A
  special focus for our future work will be the handling of
  heterogenous platforms which include dedicated hardware accelerator
  tiles that offload the regular processing elements with compute
  intensive tasks, such as crypto or signal processing accelerators.
\end{description}

%% file: debugging.tex
\section{Debugging and Diagnosis Infrastructure}

\begin{figure}
 \centering
 \includegraphics{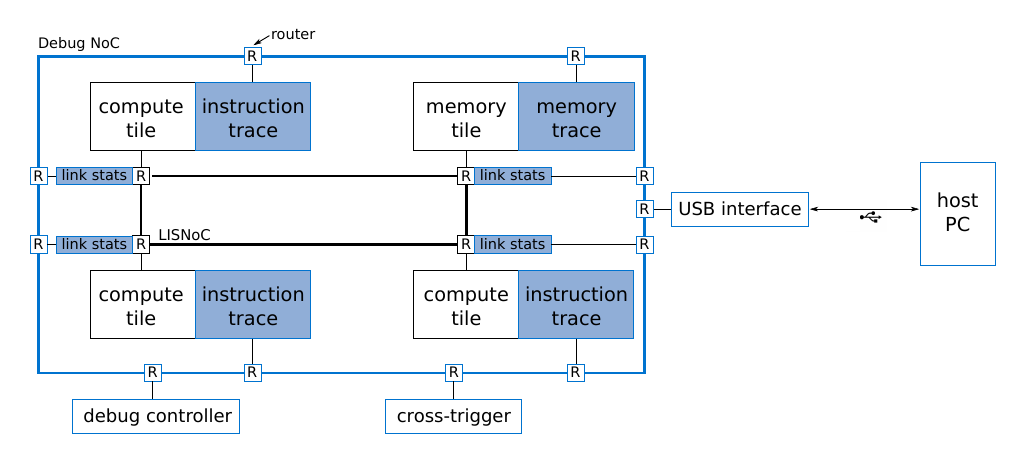}
 \caption{A sample 2$\times$2 system with the debug infrastructure
(blue).}
 \label{fig:dbg-arch-overview}
\end{figure}

Optimizing the system and writing software for OpTiMSoC, like for any other
tightly integrated System-on-Chip, is complicated by the limited
system observability. To overcome this problem a debugging and diagnosis
infrastructure was integrated into OpTiMSoC. The most significant
challenge of traditional run-control based debugging approaches, the scaling to
heterogenous multi-core platforms with different clock domains, is resolved by a
modular and decentralized tracing-based solution, as depicted in
Figure~\ref{fig:dbg-arch-overview} for an exemplary 2$\times$2 system.

Each system component that should be observed can be extended with a specific
debug module, which collects relevant data from it in the form of traces. The
content of these traces and the definition of ``relevant data'' can differ
greatly between the data sources; from CPU cores instruction traces representing
the program flow are collected, from routers in the NoC aggregated link usage
statistics, and from the memory data traces can be extracted (not yet part of
OpTiMSoC).

\medskip

Triggers can be set to reduce the amount of data collected. A trigger
is a condition that can cause the collection of trace data to start or
to stop. These conditions can also be combined across different debug
modules (cross-triggers).

All collected data is then augmented with a timestamp to enable a
temporal correlation of trace messages from different sources,
possibly compressed\footnote{Currently compression is only implemented
  for instruction traces.} and then sent over an independent, 16-bit
wide buffered Ring-NoC (the so-called Debug NoC) to an external
interface connected to a host PC.

\medskip

Depending on the target hardware different interfaces can be used to transfer
the tracing data off-chip. Currently implemented are USB~2.0 for the ZTEX~1.15
boards and TCP for OpTiMSoC running as ModelSim simulation.

On the host PC side, a generic library, liboptimsochost, abstracts
from the different transport layers and provides a high-level
interface to applications. Current user applications are optimsoc\_cli
and a basic graphical application.

%% file: roadmap.tex
\section{Project Status \& Roadmap}

A young and large project like OpTiMSoC constantly evolves in many
directions. At this time we work on integrating and documenting the
aspects covered in this paper plus some special topics, such as
support for different clock domains with the appropriate clock-domain
crossings and clock management, the debug system and other components
that have been already prepared in-house. The releases of different
elements is spread over the second quarter of 2013.

\medskip 

The focus of future work is directed towards the challenges of a shared memory
implementation (cache organization and coherence), the integration of more
hardware accelerator options, and novel approaches in the field of debugging and
diagnosis as well as in the runtime support system.

Apart from our own roadmap we hope to be able to integrate
contributions from others which might evolve OpTiMSoC in yet unknown
directions. We are glad to share our work and hope it helps other
researchers in their work and welcome any feedback or contributions.